\documentclass{emulateapj}
\usepackage{natbib}
\usepackage{apjfonts}

\usepackage{amsmath}

\begin{document}
\title{Comparing the Dark Matter Halos of Spiral, Low Surface Brightness and Dwarf Spheroidal Galaxies}
\shorttitle{Dark Matter Halos of Spirals, LSBs and dSphs}
\author{Matthew G. Walker\altaffilmark{1}, Stacy S. McGaugh\altaffilmark{2}, Mario Mateo\altaffilmark{3}, Edward W. Olszewski\altaffilmark{4} and Rachel Kuzio de Naray\altaffilmark{5,6}}
\email{walker@ast.cam.ac.uk}
\altaffiltext{1}{Institute of Astronomy, University of Cambridge, UK}
\altaffiltext{2}{Department of Astronomy, University of Maryland, College Park, MD}
\altaffiltext{3}{Department of Astronomy, University of Michigan, Ann Arbor}
\altaffiltext{4}{Steward Observatory, The University of Arizona, Tucson, AZ}
\altaffiltext{5}{Center for Cosmology, University of California, Irvine}
\altaffiltext{6}{NSF Astronomy \& Astrophysics Postdoctoral Fellow}
\begin{abstract} 
We consider dark masses measured from kinematic tracers at discrete radii in galaxies for which baryonic contributions to overall potentials are either subtracted or negligible.  Recent work indicates that rotation curves due to dark matter (DM) halos at intermediate radii in spiral galaxies are remarkably similar, with a mean rotation curve given by $\log_{10}[V_{c,\mathrm{DM}}/(\mathrm{km s^{-1}})]=1.47_{-0.19}^{+0.15}+0.5\log_{10}[r/\mathrm{kpc}]$.  Independent studies show that while estimates of the dark mass of a given dwarf spheroidal (dSph) galaxy are robust only near the half-light radius, data from the Milky Way's (MW's) dSph satellites are consistent with a narrow range of mass profiles.  Here we combine published constraints on the dark halo masses of spirals and dSphs and include available measurements of low surface brightness galaxies for additional comparison.  We find that most measured MW dSphs lie on the extrapolation of the mean rotation curve due to DM in spirals.  The union of MW-dSph and spiral data appears to follow a mass-radius relation of the form $M_{\mathrm{DM}}(r)/M_{\odot}=200_{-120}^{+200}(r/\mathrm{pc})^2$, or equivalently a constant acceleration $g_{\mathrm{DM}}=3_{-2}^{+3}\times 10^{-9}\mathrm{cm s^{-2}}$, spanning $0.02\la r \la 75$ kpc.  Evaluation at specific radii immediately generates two results from the recent literature: a common mass for MW dSphs at fixed radius and a constant DM central surface density for galaxies ranging from MW dSphs to spirals.  However, recent kinematic measurements indicate that M31's dSph satellites are systematically less massive than MW dSphs of similar size.  Such deviations from what is otherwise a surprisingly uniform halo relation presumably hold clues to individual formation and evolutionary histories. 
\end{abstract}


\section{Introduction}

Considered within the framework of Newtonian mechanics, the observed motions of stars and gas imply that the luminous components of galaxies are embedded within extended halos of unseen material.  Absent the ability to measure invisible masses directly, for a given galaxy one measures dynamical and luminous masses independently and then identifies dark matter simply as the difference between the two: $M_{\mathrm{DM}}=M_{\mathrm{dyn}}-M_{\mathrm{lum}}$.  Here we shall investigate the properties of dark matter halos only insofar as they are accessible via the direct application of this equation, without adopting any particular halo model.  

This task is made difficult by the fact that there are no galaxies for which both $M_{\mathrm{dyn}}$ and $M_{\mathrm{lum}}$ are easily measured.  While the circular motions of stars and gas in spiral galaxies provide a relatively clean measure of the dynamical mass profile, subtraction of the luminous contribution suffers from systematic uncertainties, primarily those related to stellar mass-to-light ratios.  One sidesteps this problem, or at least exchanges it for others, by considering dwarf spheroidal (dSph) galaxies, for which baryonic components contribute negligibly to internal gravitational potentials.  However, because dSphs are supported by stellar velocity dispersions instead of rotation, estimation of their individual mass profiles requires large kinematic data sets and typically employs strong modeling assumptions (see discussion by, e.g., \citealt{pryor90,wilkinson02,gilmore07}).  Larger elliptical galaxies (see \citealt{napolitano10}) combine the most challenging aspects of both regimes, as their significant stellar masses are supported in large part by random motions.

Nevertheless, recent work indicates that dark masses are constrained reasonably well at intermediate radii of both spiral and dSph galaxies.  \citet[``M07'']{mcgaugh07} use rotation curve data for a sample of 60 spirals and discard data points interior to $r<1$ kpc, where dynamical masses can be affected by noncircular motions.  After subtracting baryonic masses, M07 find that the rotation curves due exclusively to dark matter halos lie nearly on top of each other, with mean rotation curve
\begin{equation}
  \log_{10} [V_{c,\mathrm{DM}}/(\mathrm{km s^{-1}})]=1.47_{-0.19}^{+0.15}+0.5\log_{10} [r/\mathrm{kpc}].
  \label{eq:rotation}
\end{equation}

Working at smaller scales, \citet[``W09'']{walker09d} use kinematic data for eight Milky Way (MW) dSphs to show that for a wide range of plausible halo shapes and velocity anisotropies, the estimated mass within the projected half-light radius is approximately (subject to validity of the assumptions of spherical symmetry, dynamical equilibrium, flat velocity dispersion profiles $\langle v^2\rangle^{1/2}(R)=\sigma$, and negligible contributions to measured velocity dispersions from unresolved binary motions)
\begin{equation}
  M(r_{\mathrm{half}})\approx \frac{5r_{\mathrm{half}}\sigma^2}{2G}.
  \label{eq:w09}
\end{equation}
Subsequently \citet{wolf09} have provided an analytic argument for why such an estimate is insensitive to anisotropy.  Since dSph kinematics tend to be dominated by dark matter at all radii (Section \ref{subsec:dsphs}; see also \citealt{mateo98} and references therein; \citealt{walker07b}), Equation \ref{eq:w09} provides a crude estimate of the dark mass within $r_{\mathrm{half}}$ for any dSph for which measurements of $r_{\mathrm{half}}$ and $\sigma$ are available.  W09 apply Equation \ref{eq:w09} to data for 28 classical dSphs and ultrafaint satellites in the Local Group and find that, allowing for scatter by a factor of less than two in normalization, the ensemble of dSph data is consistent with a mass profile of the form $M\propto r^{1.4\pm 0.4}$.

Thus M07 and W09 independently note the apparent self-similarity of spiral and dSph dark matter halo profiles, respectively.  Here we combine data from those studies in order to examine whether the mean rotation curve attributable to dark matter in spiral galaxies extends to the small radii characteristic of dSphs.  For further comparison we also include recent measurements by \citet{kuzio08} of rotation curves in low surface brightness (LSB) galaxies.  The combined data set samples dark matter halos over the range $0.2 \la r \la 75$ kpc.  

\section{Data}

\subsection{Spirals}
\label{subsec:spirals}

The M07 data consist of HI rotation curves compiled by \citet[observational references therein]{sanders02}, trimmed for quality to a formal accuracy of $5\%$ or better \citep{mcgaugh05}, and restricted to radii larger than $r\geq 1$ kpc because systematic uncertainties (e.g., noncircular motions, beam smearing) dominate the deconvolution of rotation curves at smaller radii.  The final sample contains $686$ independent, resolved rotation velocity measurements for 60 galaxies, spanning radii $1\leq r\leq 75$ kpc.  This sample covers virtually the entire range of spiral properties, including circular velocities $50 \leq V_f \leq 300$ km s$^{-1}$, scale radii $0.5 \leq r_d\leq 13$ kpc, baryonic masses $3\times 10^8 < M_b < 4\times 10^{11}M_{\odot}$, central surface brightnesses $19.6\leq \mu_{0,b}\leq 24.2$ mag arcsec$^{2}$, and gas fractions $0.07\leq f_g\leq 0.95$.  

In order to isolate the contribution of dark matter to an observed rotation curve, M07 subtract contributions from stellar (bulge, disk) and HI components.  For this task M07 use GIPSY software \citep{vanderhulst92} to convert numerically the observed luminous distributions into rotation curves.  The most critical step in this procedure is the estimation of the stellar mass-to-light ratio, $\Upsilon_*\equiv M_*/L_*$.  For each spiral, M07 adopt the value of $\Upsilon_*$ obtained by fitting the rotation curve with modified Newtonian dynamics (MoND; \citealt{milgrom83,sanders02}).  MoND generally provides good fits to the rotation curves of spirals with $\Upsilon_*$ as the only free parameter, and the resulting estimates of $\Upsilon_*$ stand in excellent agreement with population synthesis models \citep{bell01,mcgaugh04}.  Note that despite this use of MoND as an empirically justified fitting formula with which to estimate $\Upsilon_*$, the ensuing dynamical analysis of M07 is purely Newtonian.  Alternative prescriptions for estimating $\Upsilon_*$ (e.g., maximum-disk or the population-synthesis models of \citealt{bell03}) give similar results for the dark matter rotation curves, albeit with more scatter (M07).  

\subsection{Low Surface Brightness Galaxies}

\citet{kuzio06,kuzio08,kuzio09} report high resolution observations of LSB disk galaxies.  These data are two-dimensional velocity fields of the inner regions of LSBs, augmented by long-slit and, at larger radii, HI data.  Here we include data for the nine LSBs that have mass models presented in the second part of Table 3 from \citet{kuzio08}.  LSBs are dark-matter dominated down to small radii, but their stellar components are not completely negligible in the inner kpc.  For consistency with the spiral data, we adopt the MoND estimates of $\Upsilon_*$ from \citet{deblok98} where available.  These are not available in the cases of NGC 4395, UGC 4325, and DDO 64, for which we adopt population synthesis estimates given by \citet{kuzio08}.  The MoND estimates of $\Upsilon_*$ are consistent with those of population synthesis models \citep{mcgaugh04} with most of the velocity attributed to dark matter in either case, so it makes little difference which is adopted.  The observed velocities have also been corrected for pressure support (typically $\sim 8\;\mathrm{km}\,\mathrm{s}^{-1}$); this correction is barely perceptible in most cases.  We relax the accuracy requirement to $30\%$ in velocity for the LSBs in order to retain a reasonable amount of data.  

\subsection{Dwarf Spheroidals}
\label{subsec:dsphs}

We adopt the dSph data listed in Table 1 of W09 (references to original studies therein; see erratum for updated values of $r_{\mathrm{half}}$).  These data include projected half-light radii, global velocity dispersions and luminosities measured for 28 objects in the Local Group, including satellites of the MW and M31, and two Local Group dSphs---Cetus and Tucana---that are unassociated with any obvious parent system.  To these data we add measurements for eight (six previously unmeasured) M31 satellites, adopting velocity dispersions measured by \citet{kalirai09} for AndI, AndII\footnote{For AndII, the new measurement of $\sigma=7.3\pm 0.8\mathrm{km s^{-1}}$ \citep{kalirai09} replaces the value $\sigma=9.3\pm 2.7\mathrm{km s^{-1}}$ \citep{cote99} listed in Table 1 of W09.}, AndIII, AndVII, AndX and AndXIV, and velocity dispersions measured by \citet{collins09} for And IX\footnote{For AndIX, the new measurement of $\sigma=3.0_{-3.0}^{+2.8}$ km s$^{-1}$ \citep{collins09} replaces the value $\sigma=6.8\pm 2.5 \mathrm{km s^{-1}}$ \citep{chapman05} listed in Table 1 of W09.} and AndXII. 

For each of the 34 objects in the final sample, we apply Equation \ref{eq:w09} to estimate the dynamical mass, $M(r_{\mathrm{half}})$, enclosed within the sphere having radius equal to the projected half-light radius.  We adopt for all dSphs a crude stellar mass-to-light ratio of $\Upsilon_*=1.5\pm 1.0 M_{\odot}/L_{V,\odot}$ (e.g., \citealt{martin08}).  The dark mass is then\footnote{For the Plummer surface brightness profiles adopted by W09, division of the total luminosity by $2^{3/2}$ gives the luminosity inside the sphere with radius equal to the projected half-light radius $r_{\mathrm{half}}$.} 
\begin{equation}
  M_{\mathrm{DM}}(r_{\mathrm{half}})=M(r_{\mathrm{half}})-\Upsilon_{*}\frac{L_{V}}{2^{3/2}}.  
  \label{eq:dsphdm}
\end{equation}
For comparison with the dark masses in spirals we explicitly consider for dSphs the values of $M_{\mathrm{DM}}(r_{\mathrm{half}})$ obtained from Equation \ref{eq:dsphdm}, but because most dSphs are dominated by dark matter even at their centers, these values generally are indistinguishable from the dynamical masses $M(r_{\mathrm{half}})$ estimated by W09.

We note that while the velocity dispersions of most dSphs are measured from luminous red giants, the published dispersions of five ultrafaint MW satellites---BooI, Coma, Segue-1, UMaII and Willman-1---are measured primarily from stars near the main-sequence turnoffs of these objects \citep{simon07,martin07,geha09}.  Since these stars have not yet expanded to become red giants, they can still exist as members of unresolved binary systems with separations and speeds that are smaller and faster, respectively, than those available to systems involving red giants.  Because errorbars on the velocity dispersions of these objects do not include systematic errors due to the unknown level of contamination by binary orbital motions, masses currently estimated for the faintest satellites should be considered as upper limits.  See \citet{edo96,hargreaves96b,minor10} for relevant discussions of the effects of binaries on estimates of dSph velocity dispersions.

\section{Comparing the Dark Matter Halos of Spirals, LSBs and dSphs}
\label{sec:profile}

\begin{figure}[ht]
  \plotone{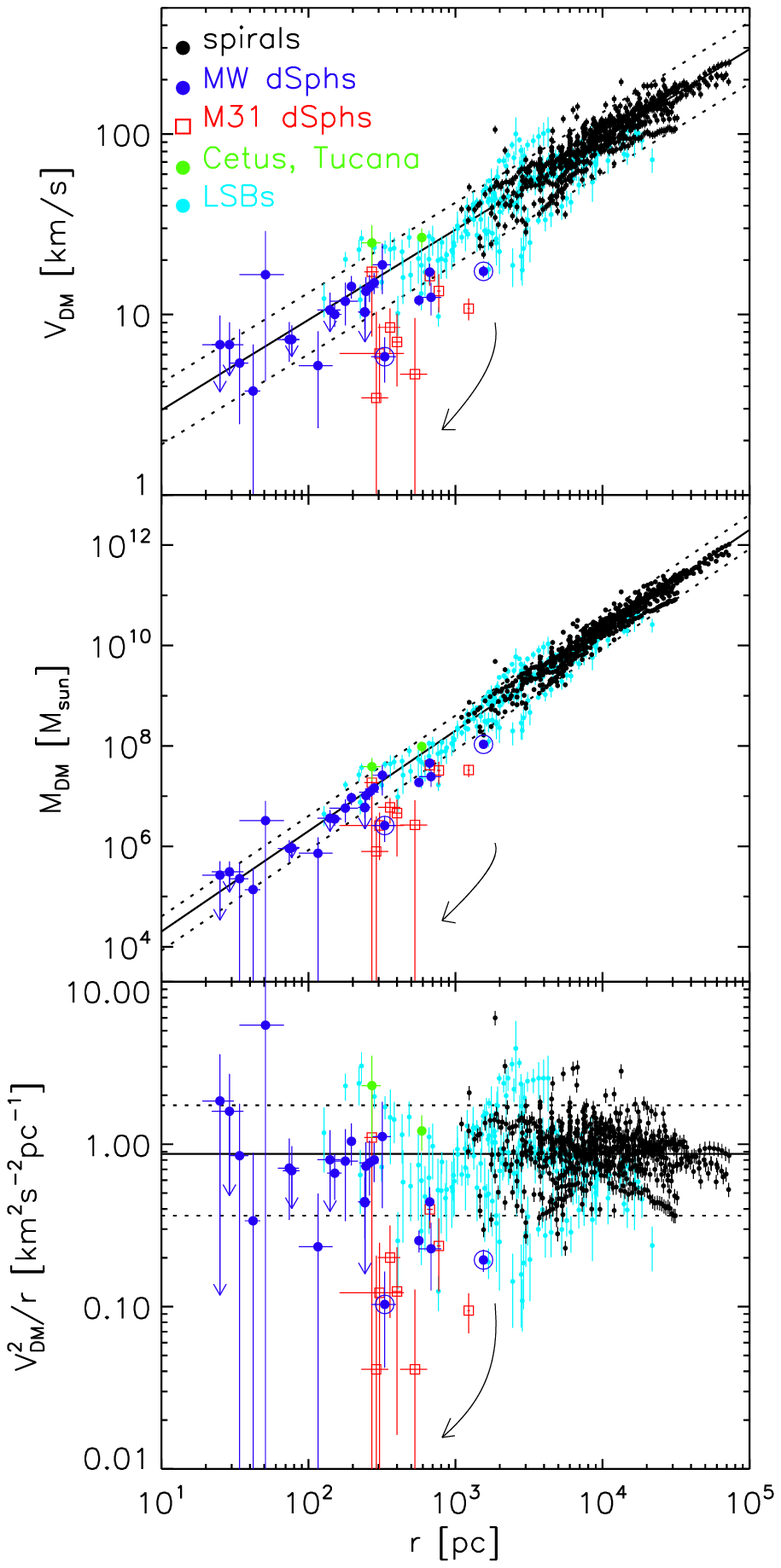}
  \caption{\scriptsize \textit{Top:} Circular velocities due to dark matter halos at intermediate radii of spirals (data from \citealt{sanders02,mcgaugh05,mcgaugh07}), LSBs (data from \citealt{kuzio06,kuzio08,kuzio09}) and within the projected half-light radii of dwarf spheroidal galaxies (data from \citealt{walker09d} and references therein; \citealt{kalirai09,collins09}).  Downward-pointing arrows indicate five ultrafaint dSphs (BooI, Coma, Seg1, UMaII and Will1) with velocity dispersions measured primarily from faint turnoff stars, for which unresolved binary orbital velocities may be significant.  Middle and bottom panels give the same information in terms of enclosed masses and accelerations, where $g_{\mathrm{DM}}=GM_{\mathrm{DM}}/r^2=V_{c,\mathrm{DM}}^2/r$.  In all panels, the solid line indicates the fit by \citet{mcgaugh07} to the spiral data alone: $\log_{10}[V_{c,\mathrm{DM}}/(\mathrm{km s^{-1}})]=1.47_{-0.19}^{+0.15}+0.5\log_{10}[r/\mathrm{kpc}]$, corresponding to a mass-radius relation of the form $M_{\mathrm{DM}}(r)/M_{\odot}= 200_{-120}^{+200}(r/\mathrm{pc})^2$, or $g_{\mathrm{DM}}= 0.9_{-0.5}^{+0.9}$ km$^2$s$^{-2}$pc$^{-1}$ (dotted lines indicate the formal scatter).  Open circles identify the two most extreme outliers---Hercules ($r_{\mathrm{half}}\sim 330$ pc) and Sagittarius ($r_{\mathrm{half}}\sim 1550$ pc)---among MW dSph satellites.  Black arrows indicate the magnitude and direction that a generic dSph would be displaced due to the tidal stripping of 99\% of the original stellar mass, from N-body simulations by \citet{penarrubia08b}.  }
  \label{fig:universal}
\end{figure}

Figure \ref{fig:universal} displays the union of spiral, LSB and dSph data.  The top panel in Figure \ref{fig:universal} plots the rotation velocity due to the dark matter halo against the radius where this quantity is evaluated.  We re-emphasize that these estimates are derived independently of any particular (i.e., cored or cusped) halo model.  Data points for spirals are the same as those in Figure 3 of M07.  For the dSphs we translate estimates of $M_{\mathrm{DM}}(r_{\mathrm{half}})$ into circular velocities using $V_{c,\mathrm{DM}}^2=GM_{\mathrm{DM}}/r$.  The middle panel contains the same information cast in terms of enclosed dark mass.  Here we plot dSph masses directly and convert circular velocities for spirals and LSBs into enclosed masses.  In both panels, a solid line indicates the fit provided by M07 to the mean rotation curve of the spiral data alone, which is given by Equation \ref{eq:rotation} or equivalently by the mass-radius relation
\begin{equation}
  \frac{M_{\mathrm{DM}}(r)}{M_{\odot}}=200_{-120}^{+200} \biggl (\frac{r}{\mathrm{pc}}\biggr )^2.
  \label{eq:mass}
\end{equation}
The bottom panel of Figure \ref{fig:universal} again conveys the same information, now cast in terms of the acceleration due to the dark matter halo, $g_{\mathrm{DM}}=V_{c,\mathrm{DM}}^2/r$.  Since the M07 relation has the form $M\propto r^2$, the corresponding acceleration profile is constant:
\begin{equation}
  g_{\mathrm{DM}}=0.9_{-0.5}^{+0.9} \mathrm{km} ^{2}\mathrm{s}^{-2}\mathrm{pc}^{-1}=3_{-2}^{+3}\times 10^{-9} \mathrm{cm}\mathrm{s}^{-2}.  
  \label{eq:acceleration}
\end{equation}

In order to evaluate whether the dSph and LSB data are consistent with the extrapolation of the M07 relation at small radii, we calculate 
\begin{equation}
  \chi^2\equiv \displaystyle \sum_{i=1}^N\frac{ (\log_{10}[V_{\mathrm{DM},i}/(\mathrm{kms^{-1}})]-0.5\log_{10}[r_i/\mathrm{kpc}]-1.47)^2}{\sigma_{\log_{10}[V_{\mathrm{DM},i}/(\mathrm{kms^{-1}})]}^2+\sigma_{\mathrm{M07}}^2}, 
  \label{eq:chi2}
\end{equation}
where residuals are scaled by measurement errors convolved with the intrinsic scatter in the M07 relation.  The value of $\sigma_{\mathrm{M07}}\sim 0.15$ reported by M07 reflects the non-Gaussian distribution of spiral velocities at a given radius.  Here we adopt a smaller value of $\sigma_{\mathrm{M07}}=0.08$ so that the central $68\%$ of spiral data points lie within $\pm \sigma_{\mathrm{M07}}$ of the M07 relation.  

For the 23 MW dSphs we obtain $\chi^2/N=2.2$.  At present the most extreme outlier is Hercules, which has a small velocity dispersion---and hence a small inferred mass---relative to other MW dSphs of similar size \citep{aden09}.  At least some of the scatter within the dSph satellite populations likely results from tidal interactions with the MW and M31.  Black arrows in Figure \ref{fig:universal} indicate, according to N-body simulations by \citet{penarrubia08b}, the magnitude and direction of the track followed by a generic satellite as it loses 99\% of its original stellar mass to tidal disruption.  Tidal evolution tends to carry individual objects toward smaller inferred masses as velocity dispersion decreases more quickly than half-light radius.  Thus tides may explain the relatively small mass estimated for Hercules, whose elongated morphology (axis ratio $\sim 3:1$) may indicate ongoing tidal disruption \citep{belokurov07,coleman07,sand09}.  Lending support to this explanation is the fact that the next most severe outlier among MW satellites is the main body of Sagittarius, from which tidal arms extend nearly twice around the sky (e.g., \citealt{majewski03}).  If we discard Hercules and Sagittarius as systems that are strongly affected by tides, the 21 remaining MW dSphs are reasonably consistent with the M07 extrapolation, having $\chi^2/N=1.2$.  If we further exclude BooI, Com, Seg1, UMaII and Wil1, we obtain $\chi^2/N=1.5$.

The 182 points contributed by the nine LSBs have $\chi^2/N=2.0$, indicating larger scatter than expected from observational errors and intrinsic scatter in the M07 relation.  However, since the LSBs appear to trace the M07 relation approximately, the comparison to M31 dSphs is perhaps the most compelling.  Masses inferred from the new kinematic measurements of M31 dsphs \citep{kalirai09,collins09} imply that these objects fall systematically \textit{below} the M07 relation, with $\chi^2/N=5.3$.  Of nine measured M31 dSphs, all but AndI, AndVII and AndXV are outliers from the trend traced by spirals and MW dSphs.  This behavior follows directly from the fact that at a given half-light radius the velocity dispersions measured for M31 dSphs are systematically smaller than those of MW dSphs.  The M31 dSphs tend also to be larger than MW dSphs of similar luminosity \citep{mcconnachie06}, although this trend appears to be limited to the brightest objects \citep{kalirai09}.  These offsets in the bulk properties of the MW and M31 satellites likely reflect differences in the environments provided by their hosts and/or the accretion histories of the satellites themselves.  Whatever the reason, it is clear that in terms of the masses attributed to their dark matter halos the M31 dSphs continue to distinguish themselves from their MW counterparts. 

\section{Discussion}
\label{sec:discussion}

M07 and W09 independently identify similarities among spiral and dSph dark mass profiles, respectively.  M07 find that the rotation curves due to dark matter in spirals lie approximately on top of each other, and W09 show that the hypothesis of a ``universal'' dSph mass profile is as well-supported by kinematic data as independent claims of a common dSph mass scale (\citealt{mateo93,strigari08}, see below).  Upon joining data sets, we find that spirals and MW dSphs---excluding outliers Her and Sgr and taking published velocity dispersions of the faintest satellites at face value---trace what appears to be a single relation under which dark mass scales with radius.  This relation can be stated in terms of a rotation curve as in M07 (Equation \ref{eq:rotation}), mass profile (Equation \ref{eq:mass}), or constant acceleration (Equation \ref{eq:acceleration}).  Before discussing implications of the new data available for M31 dSphs, we first consider the apparent scaling of MW dSph and spiral dark matter halos in the context of other reported scaling relationships involving these objects.

If the M07 relation encodes a fundamental property of dark matter halos, then along with the appearance of a universal dSph mass profile (W09), it contains and generalizes two additional results reported in the recent literature.  First, \citet{strigari08} model dSph dark matter halos and find that if they evaluate mass profiles at a fixed radius of 300 pc (requiring extrapolation of the mass profiles of the smallest dSphs to radii of $\sim 10r_{\mathrm{half}}$), MW dSphs all have $M_{300}\sim 10^7 M_{\odot}$.  If we simply evaluate the M07 mass profile at 300 pc, we obtain $M_{300}=1.8_{-1.1}^{+1.8}\times 10^7 M_{\odot}$, instantly reproducing the \citet{strigari08} result.  Here it is important to note that the M07 relation does \textit{not} imply that all dSph dark matter halos actually extend to 300 pc, or to any other radius that may be larger than that of the observed tracer population.

Second, \citet{kormendy04}, \citet{spano08} and \citet{donato09} find that if they adopt cored dark matter halos with core radius $r_0$ and central density $\rho_0$, spiral rotation curves imply a small range in central halo surface density.   \citet{donato09} extend this analysis to MW dSphs and show that over 14 magnitudes in luminosity, cored dark matter halo models imply central dark matter surface densities of $r_0\rho_0=140_{-30}^{+80} M_{\odot}$ pc$^{-2}$.  The apparent universality of $r_0\rho_0$ would then imply universality of the acceleration generated by the dark matter at $r_0$, with $g_{\mathrm{DM}}(r_0)\approx 0.5 G\pi r_0\rho_0=3.2_{-1.2}^{+1.8}\times 10^{-9}$ cm s$^{-2}$ \citep{gentile09}.  This is the same acceleration implied by the M07 relation for dark matter halos at all radii (Equation \ref{eq:acceleration}).  Thus the M07 relation unites under a single scaling relation the appearances of a ``universal mass profile'' (W09) and ``common mass scale'' for MW dSphs \citep{strigari08}, and a ``universal'' central surface density of dark matter halos \citep{donato09,gentile09}.  

The new kinematic data for M31 dSphs indicate that these objects have masses systematically smaller than their MW counterparts, placing them below the extrapolation of the M07 relation.  In order to bring the M31 dSphs onto the M07 relation, either their velocity dispersions would need to be revised upward by a factor of $\sim 2.5$ on average or their half-light radii (perhaps involving revision of distances) would need to be revised downward to $\sim 0.3$ times the current estimates.  Absent a compelling reason to suppose that the present M31 data are grossly affected by systematic errors, it seems that the formation and/or evolution of the MW and M31 systems have differed sufficiently to produce measurable differences in the inferred properties of the dark matter halos of their dSph satellites.  For example, \citet{penarrubia10} use simulations to demonstrate that tidal interactions with the baryonic disk of a host galaxy reduce the masses of dSph satellites at all radii.  These simulations reproduce the observed \textit{offset} in mass between MW and M31 dSphs by invoking a more-massive disk for M31.  Note that since the degree of mass loss depends on the parent's disk mass, this scenario requires no fine tuning of the orbital distributions of MW and M31 satellites.
 
Left unexplained is the puzzling circumstance that the dSph satellites of one but not both the MW and M31 appear to follow the same dark matter halo scaling relation as spiral galaxies.  Here we have emphasized the consistency of MW dSph data with extrapolation of the M07 relation in order to generalize recent and independent claims of uniformity among these galaxies (\citealt{strigari08,donato09,gentile09}, W09).  To the extent that the new data for M31 dSphs undermine the extrapolation of the M07 relation to dSphs, they also undermine each of these previous claims of uniformity.  On the other hand, given the susceptibility of the smallest dark matter halos to evolution driven by environment, it is feasible that one of the Local Group satellite populations has evolved significantly more than the other, altering what may have been more similar conditions initially.  In any case, it is likely that the emerging contrast between MW and M31 dSphs relates important details of the processes that shaped these two populations.  

This work follows directly from conversations that took place at the workshop \textit{Extreme Star Formation in Dwarf Galaxies} (Ann Arbor, MI, July 2009).  We are grateful to O.\ Gnedin for organizing the workshop, and to J. Pe\~narrubia, S. Koposov and J.\ Wolf for helpful discussions.  MGW acknowledges support from the STFC-funded Galaxy Formation and Evolution program at the Institute of Astronomy, Cambridge.  SSM acknowledges support from NSF grant AST-0908370.  MM and EWO acknowledge support from NSF grants AST-0808043 and AST-0807498, respectively.


\begin{thebibliography}{45}
\expandafter\ifx\csname natexlab\endcsname\relax\def\natexlab#1{#1}\fi

\bibitem[{{Ad{\'e}n} {et~al.}(2009){Ad{\'e}n}, {Wilkinson}, {Read}, {Feltzing},
  {Koch}, {Gilmore}, {Grebel}, \& {Lundstr{\"o}m}}]{aden09}
{Ad{\'e}n}, D., {Wilkinson}, M.~I., {Read}, J.~I., {Feltzing}, S., {Koch}, A.,
  {Gilmore}, G.~F., {Grebel}, E.~K., \& {Lundstr{\"o}m}, I. 2009, \apjl, 706,
  L150

\bibitem[{{Bell} \& {de Jong}(2001)}]{bell01}
{Bell}, E.~F., \& {de Jong}, R.~S. 2001, \apj, 550, 212

\bibitem[{{Bell} {et~al.}(2003){Bell}, {McIntosh}, {Katz}, \&
  {Weinberg}}]{bell03}
{Bell}, E.~F., {McIntosh}, D.~H., {Katz}, N., \& {Weinberg}, M.~D. 2003, \apjs,
  149, 289

\bibitem[{{Belokurov et al.}(2007)}]{belokurov07}
{Belokurov et al.} 2007, \apj, 654, 897

\bibitem[{{Chapman} {et~al.}(2005){Chapman}, {Ibata}, {Lewis}, {Ferguson},
  {Irwin}, {McConnachie}, \& {Tanvir}}]{chapman05}
{Chapman}, S.~C., {Ibata}, R., {Lewis}, G.~F., {Ferguson}, A.~M.~N., {Irwin},
  M., {McConnachie}, A., \& {Tanvir}, N. 2005, \apjl, 632, L87

\bibitem[{{Coleman et al.}(2007)}]{coleman07}
{Coleman et al.} 2007, \apjl, 668, L43

\bibitem[{{Collins et al.}(2009)}]{collins09}
{Collins et al.} 2009, ArXiv:0911.1365

\bibitem[{{C{\^o}t{\'e}} {et~al.}(1999){C{\^o}t{\'e}}, {Mateo}, {Olszewski}, \&
  {Cook}}]{cote99}
{C{\^o}t{\'e}}, P., {Mateo}, M., {Olszewski}, E.~W., \& {Cook}, K.~H. 1999,
  \apj, 526, 147

\bibitem[{{de Blok} \& {McGaugh}(1998)}]{deblok98}
{de Blok}, W.~J.~G., \& {McGaugh}, S.~S. 1998, \apj, 508, 132

\bibitem[{{Donato} {et~al.}(2009){Donato}, {Gentile}, {Salucci}, {Frigerio
  Martins}, {Wilkinson}, {Gilmore}, {Grebel}, {Koch}, \& {Wyse}}]{donato09}
{Donato}, F., {Gentile}, G., {Salucci}, P., {Frigerio Martins}, C.,
  {Wilkinson}, M.~I., {Gilmore}, G., {Grebel}, E.~K., {Koch}, A., \& {Wyse}, R.
  2009, \mnras, 397, 1169

\bibitem[{{Geha} {et~al.}(2009){Geha}, {Willman}, {Simon}, {Strigari}, {Kirby},
  {Law}, \& {Strader}}]{geha09}
{Geha}, M., {Willman}, B., {Simon}, J.~D., {Strigari}, L.~E., {Kirby}, E.~N.,
  {Law}, D.~R., \& {Strader}, J. 2009, \apj, 692, 1464

\bibitem[{{Gentile} {et~al.}(2009){Gentile}, {Famaey}, {Zhao}, \&
  {Salucci}}]{gentile09}
{Gentile}, G., {Famaey}, B., {Zhao}, H., \& {Salucci}, P. 2009, \nat, 461, 627

\bibitem[{{Gilmore} {et~al.}(2007){Gilmore}, {Wilkinson}, {Wyse}, {Kleyna},
  {Koch}, {Evans}, \& {Grebel}}]{gilmore07}
{Gilmore}, G., {Wilkinson}, M.~I., {Wyse}, R.~F.~G., {Kleyna}, J.~T., {Koch},
  A., {Evans}, N.~W., \& {Grebel}, E.~K. 2007, \apj, 663, 948

\bibitem[{{Hargreaves} {et~al.}(1996){Hargreaves}, {Gilmore}, \&
  {Annan}}]{hargreaves96b}
{Hargreaves}, J.~C., {Gilmore}, G., \& {Annan}, J.~D. 1996, \mnras, 279, 108

\bibitem[{{Kalirai et al.}(2009)}]{kalirai09}
{Kalirai et al.} 2009, ArXiv:0911.1998

\bibitem[{{Kormendy} \& {Freeman}(2004)}]{kormendy04}
{Kormendy}, J., \& {Freeman}, K.~C. 2004, in IAU Symposium, Vol. 220, Dark
  Matter in Galaxies, ed. {S.~Ryder, D.~Pisano, M.~Walker, \& K.~Freeman},
  377--+

\bibitem[{{Kuzio de Naray} {et~al.}(2008){Kuzio de Naray}, {McGaugh}, \& {de
  Blok}}]{kuzio08}
{Kuzio de Naray}, R., {McGaugh}, S.~S., \& {de Blok}, W.~J.~G. 2008, \apj, 676,
  920

\bibitem[{{Kuzio de Naray} {et~al.}(2006){Kuzio de Naray}, {McGaugh}, {de
  Blok}, \& {Bosma}}]{kuzio06}
{Kuzio de Naray}, R., {McGaugh}, S.~S., {de Blok}, W.~J.~G., \& {Bosma}, A.
  2006, \apjs, 165, 461

\bibitem[{{Kuzio de Naray} {et~al.}(2009){Kuzio de Naray}, {McGaugh}, \&
  {Mihos}}]{kuzio09}
{Kuzio de Naray}, R., {McGaugh}, S.~S., \& {Mihos}, J.~C. 2009, \apj, 692, 1321

\bibitem[{{Majewski} {et~al.}(2003){Majewski}, {Skrutskie}, {Weinberg}, \&
  {Ostheimer}}]{majewski03}
{Majewski}, S.~R., {Skrutskie}, M.~F., {Weinberg}, M.~D., \& {Ostheimer}, J.~C.
  2003, \apj, 599, 1082

\bibitem[{{Martin} {et~al.}(2008){Martin}, {de Jong}, \& {Rix}}]{martin08}
{Martin}, N.~F., {de Jong}, J.~T.~A., \& {Rix}, H.-W. 2008, \apj, 684, 1075

\bibitem[{{Martin} {et~al.}(2007){Martin}, {Ibata}, {Chapman}, {Irwin}, \&
  {Lewis}}]{martin07}
{Martin}, N.~F., {Ibata}, R.~A., {Chapman}, S.~C., {Irwin}, M., \& {Lewis},
  G.~F. 2007, \mnras, 380, 281

\bibitem[{{Mateo} {et~al.}(1993){Mateo}, {Olszewski}, {Pryor}, {Welch}, \&
  {Fischer}}]{mateo93}
{Mateo}, M., {Olszewski}, E.~W., {Pryor}, C., {Welch}, D.~L., \& {Fischer}, P.
  1993, \aj, 105, 510

\bibitem[{{Mateo}(1998)}]{mateo98}
{Mateo}, M.~L. 1998, \araa, 36, 435

\bibitem[{{McConnachie} \& {Irwin}(2006)}]{mcconnachie06}
{McConnachie}, A.~W., \& {Irwin}, M.~J. 2006, \mnras, 365, 1263

\bibitem[{{McGaugh}(2004)}]{mcgaugh04}
{McGaugh}, S.~S. 2004, \apj, 609, 652

\bibitem[{{McGaugh}(2005)}]{mcgaugh05}
---. 2005, \apj, 632, 859

\bibitem[{{McGaugh} {et~al.}(2007){McGaugh}, {de Blok}, {Schombert}, {Kuzio de
  Naray}, \& {Kim}}]{mcgaugh07}
{McGaugh}, S.~S., {de Blok}, W.~J.~G., {Schombert}, J.~M., {Kuzio de Naray},
  R., \& {Kim}, J.~H. 2007, \apj, 659, 149

\bibitem[{{Milgrom}(1983)}]{milgrom83}
{Milgrom}, M. 1983, \apj, 270, 365

\bibitem[{{Minor} {et~al.}(2010){Minor}, {Martinez}, {Bullock}, {Kaplinghat},
  \& {Trainor}}]{minor10}
{Minor}, Q.~E., {Martinez}, G., {Bullock}, J., {Kaplinghat}, M., \& {Trainor},
  R. 2010, ArXiv:1001.1160

\bibitem[{{Napolitano} {et~al.}(2010){Napolitano}, {Romanowsky}, \&
  {Tortora}}]{napolitano10}
{Napolitano}, N.~R., {Romanowsky}, A.~J., \& {Tortora}, C. 2010,
  ArXiv:1003.1716

\bibitem[{{Olszewski} {et~al.}(1996){Olszewski}, {Pryor}, \&
  {Armandroff}}]{edo96}
{Olszewski}, E.~W., {Pryor}, C., \& {Armandroff}, T.~E. 1996, \aj, 111, 750

\bibitem[{{Pe\~narrubia} {et~al.}(2010){Pe\~narrubia}, {Benson}, {Walker},
  {Gilmore}, {McConnachie}, \& {Mayer}}]{penarrubia10}
{Pe\~narrubia}, J., {Benson}, A.~J., {Walker}, M.~G., {Gilmore}, G.,
  {McConnachie}, A., \& {Mayer}, L. 2010, ArXiv:1002.3376

\bibitem[{{Pe{\~n}arrubia} {et~al.}(2008){Pe{\~n}arrubia}, {Navarro}, \&
  {McConnachie}}]{penarrubia08b}
{Pe{\~n}arrubia}, J., {Navarro}, J.~F., \& {McConnachie}, A.~W. 2008, \apj,
  673, 226

\bibitem[{{Pryor} \& {Kormendy}(1990)}]{pryor90}
{Pryor}, C., \& {Kormendy}, J. 1990, \aj, 100, 127

\bibitem[{{Sand} {et~al.}(2009){Sand}, {Olszewski}, {Willman}, {Zaritsky},
  {Seth}, {Harris}, {Piatek}, \& {Saha}}]{sand09}
{Sand}, D.~J., {Olszewski}, E.~W., {Willman}, B., {Zaritsky}, D., {Seth}, A.,
  {Harris}, J., {Piatek}, S., \& {Saha}, A. 2009, \apj, 704, 898

\bibitem[{{Sanders} \& {McGaugh}(2002)}]{sanders02}
{Sanders}, R.~H., \& {McGaugh}, S.~S. 2002, \araa, 40, 263

\bibitem[{{Simon} \& {Geha}(2007)}]{simon07}
{Simon}, J.~D., \& {Geha}, M. 2007, \apj, 670, 313

\bibitem[{{Spano} {et~al.}(2008){Spano}, {Marcelin}, {Amram}, {Carignan},
  {Epinat}, \& {Hernandez}}]{spano08}
{Spano}, M., {Marcelin}, M., {Amram}, P., {Carignan}, C., {Epinat}, B., \&
  {Hernandez}, O. 2008, \mnras, 383, 297

\bibitem[{{Strigari} {et~al.}(2008){Strigari}, {Bullock}, {Kaplinghat},
  {Simon}, {Geha}, {Willman}, \& {Walker}}]{strigari08}
{Strigari}, L.~E., {Bullock}, J.~S., {Kaplinghat}, M., {Simon}, J.~D., {Geha},
  M., {Willman}, B., \& {Walker}, M.~G. 2008, \nat, 454, 1096

\bibitem[{{van der Hulst} {et~al.}(1992){van der Hulst}, {Terlouw}, {Begeman},
  {Zwitser}, \& {Roelfsema}}]{vanderhulst92}
{van der Hulst}, J.~M., {Terlouw}, J.~P., {Begeman}, K.~G., {Zwitser}, W., \&
  {Roelfsema}, P.~R. 1992, in Astronomical Society of the Pacific Conference
  Series, Vol.~25, Astronomical Data Analysis Software and Systems I, ed.
  {D.~M.~Worrall, C.~Biemesderfer, \& J.~Barnes}, 131--+

\bibitem[{{Walker} {et~al.}(2007){Walker}, {Mateo}, {Olszewski}, {Gnedin},
  {Wang}, {Sen}, \& {Woodroofe}}]{walker07b}
{Walker}, M.~G., {Mateo}, M., {Olszewski}, E.~W., {Gnedin}, O.~Y., {Wang}, X.,
  {Sen}, B., \& {Woodroofe}, M. 2007, \apjl, 667, L53

\bibitem[{{Walker} {et~al.}(2009){Walker}, {Mateo}, {Olszewski},
  {Pe{\~n}arrubia}, {Wyn Evans}, \& {Gilmore}}]{walker09d}
{Walker}, M.~G., {Mateo}, M., {Olszewski}, E.~W., {Pe{\~n}arrubia}, J., {Wyn
  Evans}, N., \& {Gilmore}, G. 2009, \apj, 704, 1274

\bibitem[{{Wilkinson} {et~al.}(2002){Wilkinson}, {Kleyna}, {Evans}, \&
  {Gilmore}}]{wilkinson02}
{Wilkinson}, M.~I., {Kleyna}, J., {Evans}, N.~W., \& {Gilmore}, G. 2002,
  \mnras, 330, 778

\bibitem[{{Wolf} {et~al.}(2009){Wolf}, {Martinez}, {Bullock}, {Kaplinghat},
  {Geha}, {Munoz}, {Simon}, \& {Avedo}}]{wolf09}
{Wolf}, J., {Martinez}, G.~D., {Bullock}, J.~S., {Kaplinghat}, M., {Geha}, M.,
  {Munoz}, R.~R., {Simon}, J.~D., \& {Avedo}, F.~F. 2009, ArXiv:0908.2995

\end{thebibliography}
\end{document}